---------------------------------------------------------------------
\documentstyle[preprint,aps]{revtex}
\newcommand{\be}{\begin{eqnarray}}
\newcommand{\ee}{\end{eqnarray}}
\begin{document}
\draft
\preprint{KAIST--CHEP--94/07}
\title{ Tests of the Parton Model for Inclusive Semileptonic
B Decays with the Heavy Quark Effective Theory }
\author{
Kang Young Lee\thanks{kylee@chep5.kaist.ac.kr}
Yeong Gyun Kim\thanks{ygkim@chep6.kaist.ac.kr}
and Jae Kwan Kim
}
\vspace{.5in}
\address{
Dept. of Physics, KAIST, Taejon 305-701, KOREA
}
\date{\today}
\maketitle
\begin{abstract}
In order to study the end-point spectrum of inclusive semileptonic
$b \rightarrow u$ decays, we investigate the parton model motivated by
deep inelastic scattering process. The reliability of the model is carefully
examined by comparing the decay rates and the lepton spectra
for $b \rightarrow c$ decay with those of the heavy quark effective theory.
We also compute the decay rates of the inclusive semileptonic decay into
tau lepton in terms of the parton model, which give another
test of the model independently.
We conclude that one can make the parton model consistent with
the heavy quark effective theory, and it would be
phenomenologically useful in extracting $|V_{ub}|$.
\\
\end{abstract}
\pacs{ }

\narrowtext


Semileptonic decays of B mesons play an important role for understanding
the Standard Model (SM). They are essential for testing and determining
the parameters in the SM, and also provide valuable informations on
the bound state structure of the mesons.

Recently, there has been much progress in the study of inclusive
semileptonic decays of $b$ hadrons with the help of heavy quark effective
theory (HQET) \cite{heavy,blok,manohar}. One can get a QCD-based expansion
in powers of $1/m_Q$ by performing an operater product expansion
on the hadronic tensor.
It has been shown that at leading order the contributions to the decay rates
in the expansion coincide with those of the free quark decay.
The terms of order of $1/m^2_Q$ are the leading corrections of
the nonperturbative effects because terms of
order of $1/m_Q$ vanish.
These enable one to obtain model-independent calculations
of the decay rates and lepton spectra
for most region of phase space.

However, a model study is still required.
The study of the end-point region of the electron spectrum
in $b \rightarrow u$ decay where $\bar{B} \rightarrow X_c l \bar{\nu}$
decays are forbidden
is of great interest in the decay of $\bar{B} \rightarrow X_u l \bar{\nu}$,
because it is potentially useful to determine
the Cabbibo-Kobayashi-Maskawa (CKM) angle $|V_{ub}|$.
Unfortunately, the HQET calculation of the
lepton spectrum for the inclusive semileptonic decay gives
singularities in the end-point region \cite{blok,manohar}
and modelling the end-point is inevitable
for the extraction of $|V_{ub}|$ at present.
Thus the following strategy may be
presented: one picks a model, fixes its shape parameters
from fitting the spectrum in $\bar{B} \rightarrow X_c l \bar{\nu}$
and applies it to the end-point region from where one extracts
$|V_{ub}/V_{cb}|$.
Up to now the most popular model for the inclusive semileptonic decays of
heavy mesons is the model presented by Altarelli et al. \cite{accmm}, to be
refered as ACCMM. They improve the free quark decay by considering
the Fermi motion of the spectater quark in the meson. The ACCMM model
has been reanalysed in the framework of the HQET recently by several authors
\cite{accmmh}.
They showed that the ACCMM model can be made to be consistent with
the HQET by properly redefining the model.

In the recent publication \cite{parton1}, a parton model motivated
by the deep inelastic scattering process (DIS) was developed
and the results of the model was compared with experimental results.
It is well known that inclusive semileptonic decays of heavy flavours
are intimately related to DIS via channel crossing.
The model shows a good agreement with the experimental data and
yields well-behaved lepton energy spectrum
at the end-point region where it may serve as a useful model.
The parton model seems to be one of the
useful models that can describe this region.
In this letter, we investigate the
reliability of the parton model approach by explicitly comparing
the results of the model with those of HQET.
For the model to be phenomenologically reliable, we require that
the decay rates and the lepton spectra
(at least far from the end-point region)
coincide with those of the HQET.
This requirement would fix the parameter of the model and
we can compare the lepton spectra of the parton model
with those of the HQET.
For the test of consistency of the model,
we also study the lepton mass effect.
Theoretically the lepton mass does not affect the picture of the parton model
and we expect that the parameters which we have fixed in the
case of the decay into an electron as a final state
should be also valid in the decay into a massive lepton.

The lepton mass effects are significant in the semileptonic decay
involving a tau lepton.
The tau channel in semileptonic decays of B mesons is one of the
latest measured process. Recently a branching ratio
Br($\bar{B} \rightarrow X \tau \bar{\nu}$) has been measured in LEP \cite{L3}.
\be
\mbox{Br}(\bar{B} \rightarrow X \tau \bar{\nu}) = 2.4 \pm 0.7~(\mbox{stat.})
                                               \pm 0.8 ~(\mbox{syst.})~\%
\nonumber
\ee
This mode is of some theoretical interest by several reasons,
since the decay kinematics and dynamics become complicated
in the decay involving a tau lepton.
There are more structure functions which does not contribute to
the decays involving an electron. Since these changes are independent of
the CKM angle $|V_{cb}|$ and $|V_{ub}|$, computation of the ratio
of the decay rates for the tau lepton channel to those for the electron
channel is useful in testing the model. Phenomenologically
the study of $B \rightarrow X \tau \bar{\nu}$ is useful
to determine the CKM angle $|V_{cb}|$
independently of the decay into an electron.
This channel is also interesting as being the window to show
the effects of physics beyond the SM.
It has been used to probe the charged Higgs boson signal and gives some
constraints on parameters of the minimal supersymmetric
standard model \cite{isidori}.

In the following we briefly review the parton model for inclusive
semileptonic decays of B mesons \cite{parton1,parton2}
and describe its extension to the decay of $B \rightarrow X \tau \bar{\nu}$.
The main ideas of this model are followings: $i)$
it pictures the mesonic decay as the decay of the partons. We
consider that the parton in the B meson carries a fraction
of the B meson momentum. $ii)$ The distribution of a parton in the
B meson is determined by a single function, which will be discussed
later.

The hadronic tensor is defined as
\be
4\pi W^{\mu \nu} &=& \int d^4x ~~e^{iqx} <B|J^{\mu}(x) J^{\nu}(x)|B>
                 \nonumber \\
                 &=& \int \frac{d^3 P_X}{(2\pi)^3 2E_X}
                          (2\pi)^4 \delta^4(P_B-P_X-q) {\tilde W}^{\mu \nu},
\ee
where $P_X$ are the momentum of the final state hadrons.
Using standard definitions of the structure functions
\be
W^{\mu \nu} &=& -g^{\mu \nu} W_1 + P_B^{\mu} P_B^{\nu} \frac{W_2}{m_B^2}
                -i \epsilon^{\mu \nu \alpha \beta} {P_B}_{\alpha} q_{\beta}
                \frac{W_3}{2 m_B^2} \nonumber \\
             && + q^{\mu} q^{\nu} \frac{W_4}{m_B^2}
                +(P_B^{\mu} q^{\nu} + P_B^{\nu} q^{\mu}) \frac{W_5}{2 m_B^2}
                +i(P_B^{\mu} q^{\nu} - P_B^{\nu} q^{\mu}) \frac{W_6}{2 m_B^2},
\ee
we obtain
\be
W_1 &=& \frac{1}{2} \left(f(x_{+})-f(x_{-})\right), \nonumber \\
\frac{W_2}{m_B^2} &=& \frac{2}{m_B^2 (x_+-x_-)}
                     \left( x_{+}f(x_{+}) + x_{-}f(x_{-}) \right),
                     \nonumber \\
\frac{W_3}{2 m_B^2} &=& -\frac{1}{m_B^2 (x_+-x_-)}
                     \left( f(x_{+}) + f(x_{-}) \right),
                     \nonumber \\
\frac{W_5}{2 m_B^2} &=& -\frac{1}{m_B^2 (x_+-x_-)}
                     \left( f(x_{+}) + f(x_{-}) \right),
                     \nonumber \\
\ee
where
\be
x_{\pm} = \frac{q_0 \pm \sqrt{p^2+m_1^2}}{m_B}.
\ee
The kinematics give the momentum transfer $q = P_B-P_X$ which satisfies
the following relations in the B-rest frame
\be
2m_Bq_0 &=& m_B^2-m_X^2+q^2, \nonumber \\
2m_B p  &=& (m_B^4+m_X^4+q^4-2 m_B^2 m_X^2-2 m_B^2 q^2-2 m_X^2q^2)
            ^{\frac{1}{2}},
\ee
where $q=(q_0,{\bf p})$ and $p=|{\bf p}|$.
$f(x)$ is the distribution function of b quark in the B mesons and $m_1$
a final state parton mass and $m_X$ an invariant mass of the
final state hadrons. In the integrations the Dirac delta function
has two positive roots $x_+$, $x_-$ because the momentum transfer
$q^2 > 0$ in the decay process.
On the other hand, in the deep inelastic scattering process
where $q^2 < 0$
the Dirac delta function has only one positive root.
In the eq. ( 3 ) the structure function $W_5$ contributes
when the final lepton mass is nonvanishing.

The b quark distribution inside B mesons is related to the fragmentation
function by crossing symmetry of the light quark followed by a time
reversal transformation.
Thus we use the measured fragmentation
function of B mesons as the distribution function of b quark.
Specifically we take the Peterson form \cite{peterson} following the ref.
\cite{parton1,parton2},
which is the fragmentation function usually used
in Lund Monte Carlo programs:
\be
f(x) = \frac{N}{x(1-\frac{1}{x}-\frac{\epsilon_p}{1-x})^2}.
\ee
This is a single parameter function with a parameter $\epsilon_p$
and N is the normalization factor.

For the $b \rightarrow u$ decays, the decay rates depends only on a single
parameter $\epsilon_p$. Though
we need two parameters $\epsilon_p$ and $m_c$
for the $b \rightarrow c$ decays,
we may use the value of $m_c$ used in the calculation
of HQET for the results to be consistent with those of HQET.
The heavy quark mass was calculated up to $1/m_Q$ in terms of
hadronic parameters with HQET and QCD sum rules \cite{falk1,neubert}:
\be
m_{H_Q} = m_Q + {\bar \Lambda} - \frac{\lambda_1+3\lambda_2}{2 m_Q}.
\ee
The parameter $\bar{\Lambda}$ is associated with the leading
contribution to the mass
of the light degrees of freedom of the heavy meson.
The parameter $\lambda_1$ is the spin-averaged mass shift.
The parameter ${\bar \Lambda}$ and $\lambda_1$ are related to
some nonperturbative calculations. We choose the value of them
from ref. \cite{falk2}, which are calculated by QCD sum rules,
\be
0.45 ~\mbox{GeV} < &\bar{\Lambda}& < 0.60 ~\mbox{GeV},
\nonumber \\
0 ~\mbox{GeV}^2 < &-\lambda& < 0.3 ~\mbox{GeV}^2.
\ee
The value of $\lambda_2$ is obtained from the $B~-~B^*$ mass spliting
and we use the value $\lambda_2 = 0.12$ GeV$^2$.
The masses of B- and D- mesons are refered from the ref. \cite{pdb}.
\be
m_B = 5.279~~ \mbox{GeV},&& ~~~~~~m_D= 1.865~~ \mbox{GeV}\nonumber
\ee
With these values we obtain the quark masses in eq. (7) as followings:
\be
m_b = 4.776~~ \mbox{GeV},&& ~~~~~~m_c = 1.414~~ \mbox{GeV}.
\ee

Now we fix the parameters of the model using the results of HQET.
We have the total width in terms of HQET \cite{blok,manohar}
\be
\Gamma_{\mbox{HQET}} = |V_{qb}|^2 \frac{G_F^2 m_b^5}{192 \pi^3}
     \left[ (1-K_b) z_0 + G_b z_1  \right]
\ee
where $z_0=1-8\rho+8\rho^3-\rho^4-12\rho^2 \ln \rho$,
 $z_1=3-8\rho+24\rho^2-24\rho^3+5\rho^4+12\rho^2 \ln \rho$
and $\rho = m_q^2/m_b^2$. $G_b$ and $K_b$ are related to the terms
of the operator product expansion and obtained from the parameters
$\lambda_1$ and $\lambda_2$.

The parton model has the shape parameter $\epsilon_p$. We show
the decay rates of the decay $\bar{B} \rightarrow X e \bar{\nu}$
in the parton model,
$\Gamma_{\mbox{parton}}(\epsilon_p)$,
with varying $\epsilon_p$ in the table 1. In the numerical calculation
we use above values of parameters and the mass of tau lepton
\be
m_{\tau}= 1.777~~ \mbox{GeV}\nonumber
\ee
for the value to be consistent with those of ref. \cite{falk2}.
We require the condition
$ \Gamma_{\mbox{HQET}} = \Gamma_{\mbox{parton}}$ to fix the parameter
$\epsilon_p$.
 From fig. 1 we favor the values of $\epsilon_p=$ 0.0035 -- 0.004.
There has been some controversy for the value of $(m_X)_{min}$ in the decay
of b to c, whether it corresponds to $m_c$ or $m_D$.
If we measure the $m_X$ spectrum in the experiment, it has the minimun
at $m_D$ as a resonance. But if we calculate the $m_X$ spectrum
we would not obtain such features as the resonant spectra of $D$, $D^*$ etc.
in the spectrum.
It means that the $m_X$ spectrum in our calculation
is, in fact, corresponding to the momentum spectrum of
the final state parton instead of the invariant mass of the
final state hadrons. Therefore we should take the
$(m_X)_{min}$ as $m_c$.
We may compute the electron spectrum and compare it
with those of the HQET.
Figs. 2 shows normalized spectrum of the electron energy
for the parameter $\epsilon_p =$ 0.004
in both of the cases $b \rightarrow c$ and $b \rightarrow u$ along with
predictions by HQET in dotted curve.
They agrees well with those of HQET far from the end-point region.

We also compute the decay rates of
$\bar{B} \rightarrow X \tau \bar{\nu}$ in terms of the parton model..
Several authors have studied this channel in the framework
of HQET \cite{falk2,htau}. It is convenient to use the ratio
of the branching ratios of the decay with a tau lepton as a final state
to those of the decay with an electron in analyzing this channel,
because we are free from uncertainties of CKM angle
$|V_{cb}|$ and $|V_{ub}|$. Furthermore the shape parameter
$\epsilon_p$ is associated with bound state structure of B meson and must be
independent of the lepton flavours.
Therefore the results for this channel which we compute with the value of
$\epsilon_p$ obtained from the results of table 1 should also
agree with those of the HQET. Falk et al. \cite{falk2} obtain the ratio
\be
\frac{\Gamma(\bar{B} \rightarrow X \tau \bar{\nu})}
     {\Gamma(\bar{B} \rightarrow X e \bar{\nu})} = 0.215 \pm 0.035
\ee
using the HQET and our result is
\be
\frac{\Gamma(\bar{B} \rightarrow X \tau \bar{\nu})}
     {\Gamma(\bar{B} \rightarrow X e \bar{\nu})} = 0.213
\ee
They also show a good agreement with each other.
The QCD corrections are incorporated like ref.
\cite{hokim}. With the value $\alpha_s(m_b) \approx 0.22$
the total rates are corrected by the factors 0.90 for
$\bar{B} \rightarrow X \tau \bar{\nu}$ and 0.88 for
$\bar{B} \rightarrow X e \bar{\nu}$ respectively.
In the tau channel the tau lepton spectrum is not measured directly
and it is less valuable phenomenologically.

In conclusion the parton model provides a reliable method for analysing
inclusive semileptonic decays of B mesons.
The parton model has one single parameter. Even though it also
depends upon the c-quark mass, $m_c$, we may take its value from
the HQET since the results of the HQET also contain $m_c$.
We fix the value of the parameter $\epsilon_p$ by requiring the
condition $ \Gamma_{\mbox{HQET}} = \Gamma_{\mbox{parton}}$.
As independent test of our choice of the parameter, we also
calculate the electron energy spectrum and compare the results of the
model with those of HQET explicitly and it shows a good agreement with HQET.
The parton model gives well-behaved lepton spectra in the end-point
region of the electron energy by a compact formular
with a single parameter.
We also study the semileptonic B decay into tau lepton in terms of the parton
model and it provides another test of the model.
It was shown that we can consistently fix the model by investigating
three quantities, decay rates of b to c and b to u into the electron
and the ratio of the decay rates into the electron to those into
the tau lepton, which are independent of one another.
 From these we assert that the parton model is very useful in
analyzing the B decay process, especially near the end-point region
where the systematic expansion of HQET breaks down.

\acknowledgements

We would like to thank P. Ko for helpful discussions.
This work is supported in part by Korean Science and Engineering Foundation
(KOSEF).


\begin{table}
\caption{
The decay width of $\bar{B} \rightarrow X e \bar{\nu}$ computed by
the parton model with varying the parameter $\epsilon_p$.
}
\begin{center}
\begin{tabular}{cccccccccr}
&&&&&&&&&\\
& $\epsilon_p $ & & & $\Gamma(b \rightarrow c)/|V_{cb}|$ & $(\times10^{-11})$&
&
$\Gamma(b \rightarrow u)/|V_{ub}|$ & $(\times10^{-11})$&
\\
& & & &parton model& HQET & &parton model& HQET &\\
\hline
&&&&&&&&&\\
&0.0030& & &2.901& & &5.375& & \\
&0.0035& & &2.795& & &5.200& & \\
&0.0040& & &2.700&2.718& &5.042&5.189& \\
&0.0045& & &2.620& & &4.906& & \\
&0.0050& & &2.544& & &4.779& & \\
&0.0055& & &2.477& & &4.664& & \\
&0.0060& & &2.416& & &4.559& & \\
&0.0065& & &2.359& & &4.462& & \\
&0.0070& & &2.305& & &4.370& & \\
&0.0075& & &2.257& & &4.285& & \\
&0.0080& & &2.211& & &4.206& & \\
&0.0085& & &2.168& & &4.130& & \\
&0.0090& & &2.128& & &4.060& & \\
&&&&&&&&&\\
\end{tabular}
\end{center}
\end{table}

%
%

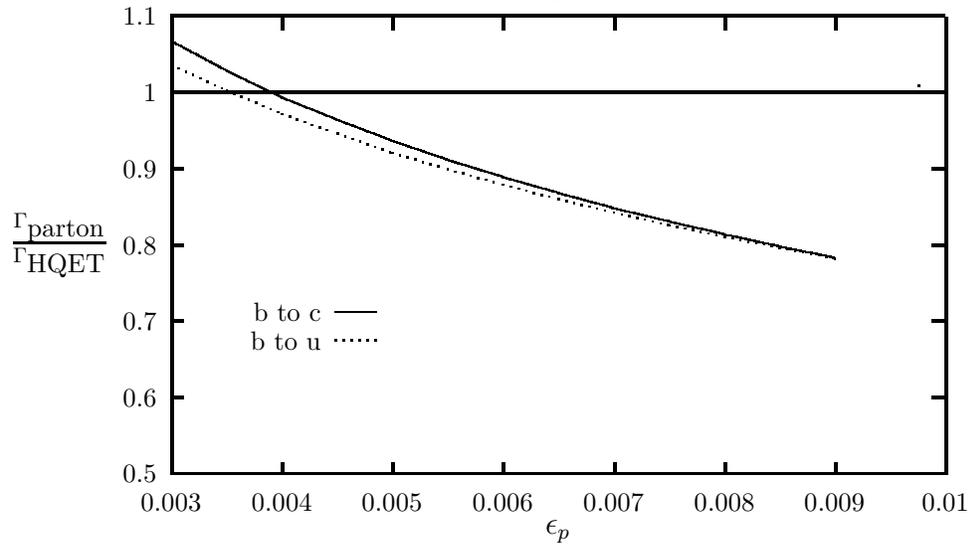
\begin{figure}
\caption{
This is the ratios of decay rates calculated in the parton model
with varying $\epsilon_p$ to those of the HQET, $\Gamma_{\mbox{parton}}
/\Gamma_{HQET}$. The solid line is of the decay $b \rightarrow c$
and the dashed line $b \rightarrow u$.
}
\label{figone}
\end{figure}
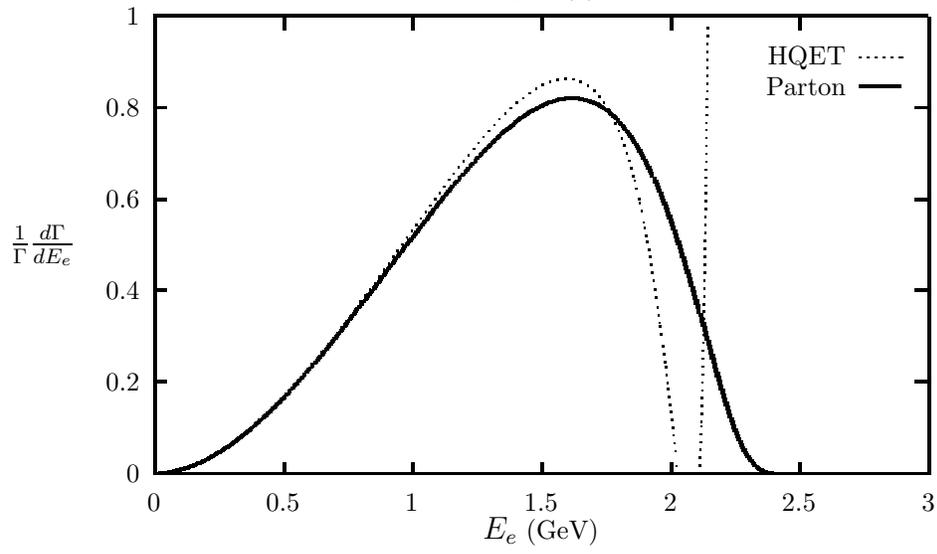
\begin{figure}
\caption{
(a) The electron energy spectra for the decays of $b \rightarrow c$.
(b) The electron energy spectra for the decays of $b \rightarrow u$.
The solid line is of the parton model and the dashed line of
the heavy quark effective theory.
The parton model is computed with the parameter $\epsilon_p=
0.004$.
}
\label{figtwo}
\end{figure}
\newpage
%
%
%
\pagestyle{empty}
\setlength{\unitlength}{0.240900pt}
\ifx\plotpoint\undefined\newsavebox{\plotpoint}\fi
\sbox{\plotpoint}{\rule[-0.500pt]{1.000pt}{1.000pt}}%

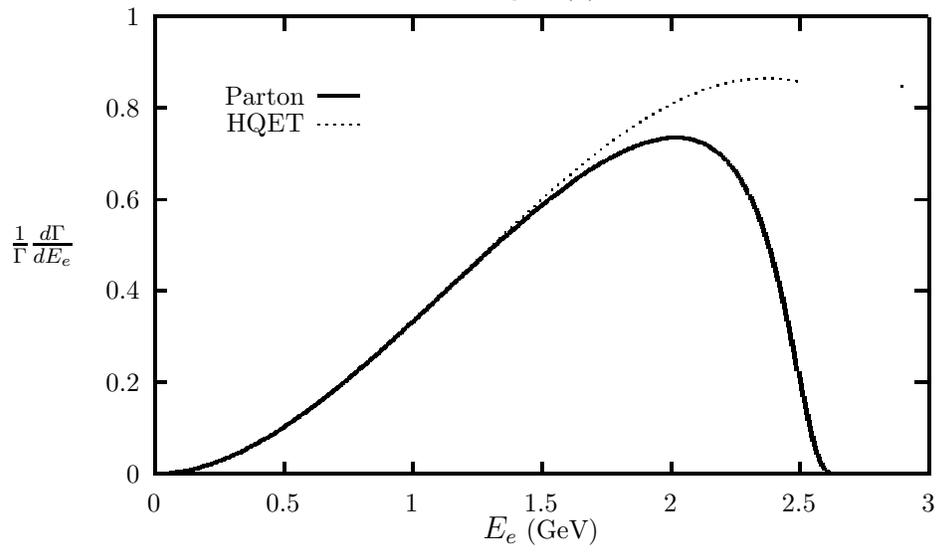
\end{document}